# Growth parameters of $Bi_{0.1}Y_{2.9}Fe_5O_{12}$ thin films for high frequency applications

Ganesh Gurjar[1,4], Vinay Sharma[2], S. Patnaik[1,*], Bijoy K. Kuanr[3]

[1]School of Physical Sciences, Jawaharlal Nehru University, New Delhi, INDIA 110067
[2]Department of Physics, Morgan State University, Baltimore, MD, USA 21251
[3]Special Centre for Nanosciences, Jawaharlal Nehru University, New Delhi, INDIA 110067
[4]Shaheed Rajguru College of Applied Sciences for Women, University of Delhi, INDIA 110096

## Abstract

The growth and characterization of Bismuth (Bi) substituted YIG (Bi-YIG, $Bi_{0.1}Y_{2.9}Fe_5O_{12}$) thin films are reported. Pulsed laser deposited (PLD) films with thicknesses ranging from 20 to 150 nm were grown on Gadolinium Gallium Garnet substrates. Two substrate orientations of (100) and (111) were considered. The enhanced distribution of $Bi^{3+}$ ions at dodecahedral site along (111) is observed to lead to an increment in lattice constant from 12.379 Å in (100) to 12.415 Å in (111) oriented films. Atomic force microscopy images showed decreasing roughness with increasing film thickness. Compared to (100) grown films, (111) oriented films showed an increase in ferromagnetic resonance linewidth and consequent increase in Gilbert damping. The lowest Gilbert damping values are found to be $(1.06\pm0.12) \times10^{-4}$ for (100) and $(2.30\pm0.36) \times10^{-4}$ for (111) oriented films with thickness of ≈150 nm. The observed values of extrinsic linewidth, effective magnetization, and anisotropic field are related to thickness of the films and substrate orientation. In addition, the in-plane angular variation established four-fold symmetry for the (100) deposited films unlike the case of (111) deposited films. This study prescribes growth conditions for PLD grown single-crystalline Bi-YIG films towards desired high frequency and magneto-optical device applications.

**Corresponding authors:** spatnaik@mail.jnu.ac.in

## 1.1    Introduction

One of the most important magnetic materials for studying high frequency magnetization dynamics is the Yttrium Iron Garnet (YIG, $Y_3Fe_5O_{12}$). Thin film form of YIG have attracted a huge attention in the field of spintronic devices due to its large spin-wave propagation length, high Curie temperature $T_c \approx 560$ K [1], lowest Gilbert damping and strong magneto-crystalline anisotropy [2-7]. Due to these merits of YIG, it finds several applications such as in magneto-optical (MO) devices, spin-caloritronics [8,9], and microwave resonators and filters [10-14].

The crystal structure of YIG is body centered cubic under $Ia\bar{3}d$ space group. In Wyckoff notation, the yttrium (Y) ions are located at the dodecahedral 24c sites, whereas the Fe ions are located at two distinct sites; octahedral 16a and tetrahedral 24d.  The oxygen ions are located in the 96h sites [7]. The ferrimagnetism of YIG is induced via a super-exchange interaction at the '$d$' and '$a$' site between the non-equivalent $Fe^{3+}$ ions. It has already been observed that substituting Bi/Ce for Y in YIG improves magneto-optical responsivity [13,15-21]. In addition, Bi substitution in YIG (Bi-YIG) is known to generate growth-induced anisotropy, therefore, perpendicular magnetic anisotropy (PMA) can be achieved in Bi doped YIG, which is beneficial in applications like magnetic memory and logic devices [7,22,23]. Due to its usage in magnon-spintronics and related disciplines such as caloritronics, the study of fundamental characteristics of Bi-YIG materials is of major current interest due to their high uniaxial anisotropy and Faraday rotation [17, 24-27]. Variations in the concentration of $Bi^{3+}$ in YIG, as well as substrate orientation and film thickness, can improve strain tuned structural properties and magneto-optic characteristics. As a result, selecting the appropriate substrate orientation and film thickness is important for identifying the growth of Bi-YIG thin films.

The structural and magnetic characteristics of Bi-YIG [$Bi_{0.1}Y_{2.9}Fe_5O_{12}$] thin film have been studied in the current study. Gadolinium Gallium Garnet (GGG) substrates with orientations of (100) and (111) were used to grow thin films. The Bi-YIG films of four different thickness (≈20 nm, 50 nm, 100 nm and150 nm) were deposited in-situ by pulsed laser deposition (PLD) method [19,28] over single-crystalline GGG substrates. Along with structural characterization of PLD grown films, magnetic properties were ascertained by using vibrating sample magnetometer (VSM) in conjunction with ferromagnetic resonance (FMR) techniques. FMR is a highly effective tool for studying magnetization dynamics. The FMR response not only provides information about the magnetization dynamics of the material such as Gilbert damping and anisotropic field, but also about the static magnetic properties such as saturation magnetization and anisotropy field.

## 1.2    Experiment

Polycrystalline YIG and Bi-YIG targets were synthesized via the solid-state reaction method. Briefly, yttrium oxide ($Y_2O_3$) and iron oxide ($Fe_2O_3$) powders from Sigma-Aldrich were grounded for ≈14 hours before calcination at 1100 °C. The calcined powders were pressed into pellets of one inch and sintered at 1300 °C. Using these polycrystalline YIG and Bi-YIG targets, thin films of four thicknesses (≈20 nm, 50 nm, 100 nm, and 150 nm) were synthesized in-situ on (100)- and (111)-oriented GGG substrates using the PLD method. The samples are labelled in the text as 20 nm (100), 20 nm (111), 50 nm (100), 50 nm (111), 100 nm (100), 100 nm (111), 150 nm (100), and 150 nm (111). Before deposition, GGG substrates were cleaned in an ultrasonic bath with acetone and isopropanol for 30 minutes. The deposition chamber was cleaned and evacuated to 5.3×10$^{-7}$ mbar. For PLD growth, a 248 nm KrF excimer laser (Laser fluence (2.3 J cm$^{-2}$) with 10 Hz pulse rate was used to ablate the material from the target. Oxygen pressure, target-to-substrate distance, and substrate temperature were maintained at 0.15 mbar, 5.0 cm, and 825 $^{\circ}$C, respectively. Growth rate of deposited films were 6 nm/min. The as-grown films were annealed in-situ for 2 hours at 825 $^{\circ}$C in the presence of oxygen (0.15 mbar). The structural characterization of thin films were ascertained using X-ray diffraction (XRD) with Cu-K$_{\alpha}$ radiation (1.5406 Å). We have performed the XRD measurement at room temperature in $\theta$-$2\theta$ geometry and incidence angle are 20 degrees. The film's surface morphology and thickness were estimated using atomic force microscopy (AFM) (WITec GmbH, Germany). The magnetic properties were studied using a vibrating sample magnetometry (VSM) in *Cryogenic* 14 Tesla Physical Property Measurement System (PPMS). FMR measurements were done on a coplanar waveguide (CPW) in a flip-chip arrangement with a dc magnetic field applied perpendicular to the high-frequency magnetic field (h$_{RF}$). A Keysight Vector Network Analyzer was used for this purpose. The CPW was rotated in the film plane from 0º to 360º for in-plane ($\phi$) measurements and from 0º to 180º for out of plane ($\theta$) measurement.

In this study, the thickness of Bi-YIG was determined by employing methods such as laser lithography and AFM. We have calibrated the thickness of thin films with PLD laser shots. Photoresist by spin coating is applied to a silicon substrate, and then straight-line patterns were drawn on the photoresist coated substrates using laser photolithography. The PLD technique was used to deposit thin films of the required material onto a pattern-drawn substrate. It is then necessary to wet etch the PLD grown thin film in order to remove the photoresist coating. Then, AFM tip is scanned over the line pattern region in order to estimate the thickness of the grown samples from the AFM profile image.

## 1.3    Results and Discussion

### 1.3.1    Structural properties

Figure 1 (a)-(d) show the XRD pattern of (100)- and (111)-oriented Bi-YIG grown thin films with thickness ≈20-150 nm (Insets depict the zoomed image of XRD patterns). XRD data indicate single-crystalline growth of Bi-YIG thin films. Figures 1 (e) and 1 (f) show the lattice constant and lattice mismatch (with respect to substrate) determined from XRD data, respectively. The cubic lattice constant $a$ is calculated using the formula,

$$a = \frac{\lambda \sqrt{h^2+k^2+l^2}}{2 \sin \theta} \qquad (1)$$

where the wavelength of Cu-K$_a$ radiation is represented by $\lambda$, diffraction angle by $\theta$, and the Miller indices of the corresponding XRD peak by [h, k, l]. Further, the lattice mismatch parameter $(\frac{\Delta a}{a})$ is calculated using the equation,

$$\frac{\Delta a}{a} = \frac{(a_{film} - a_{substrate})}{a_{film}} \times 100\% \qquad (2)$$

Here lattice constant of film and substrate are represented by $a_{film}$ and $a_{substrate}$, respectively. The reported lattice constant values are consistent with prior findings [15,17,21]. Lattice constant slightly increases with the increase in thickness of the film in the case of (111) as compared to (100). Since the distribution of Bi$^{3+}$ in the dodecahedral site is dependent on the substrate orientation [7,23,29], the (111) oriented films show an increase in the lattice constant. In Bi-YIG films, this slight increase in the lattice constant (in the 111 direction) leads to a comparatively larger lattice mismatch as seen in Fig. 1 (f). For 50 nm (111) Bi-YIG film, we achieved a lattice mismatch of ~0.47 %, which is close to what has been reported earlier [30,31].

Smaller value of lattice mismatch can reduce the damping constant of the film [31]. We want to underline the importance of lattice plane dependent growth in conjunction with film thickness in indicating structural and magnetic property changes.

### 1.3.2   Surface morphology

Figure 2 (a)-(h) shows room temperature AFM images with root mean square (RMS) roughness. Roughness is essential from an application standpoint because the roughness directly impacts the inhomogeneous linewidth broadening which leads to increase in the Gilbert damping. We have observed RMS roughness around 0.5 nm or less for all grown Bi-YIG films which are comparable to previous reported YIG films [32,33]. We have observed that RMS roughness decreases with increase in thickness of the film. With (100) and (111) orientations, there is no discernible difference in roughness. Furthermore, roughness would be more affected by changes in growth factors and by substrate orientation [7,33,34].

### 1.3.3   Static magnetization study

The room temperature ($\approx$296 K) VSM magnetization measurements were carried out with applied magnetic field parallel to the film plane (in-plane). The paramagnetic contributions from the GGG substrate were carefully subtracted. Figure 3 (a)-(h) shows the magnetization plots of Bi-YIG thin films of thickness $\approx$20-150 nm. Inset of Fig. 3 (i) shows the measured saturation magnetization ($\mu_0 M_S$) data of as-grown (100) and (111)-oriented Bi-YIG films which are consistent with the previous reports [6,17,22,35,36]. Figure 3 (i) shows plot of $\mu_0 M_s \times t$ Vs. t, where 't' is film thickness. This is done to determine thickness of dead layer via linear extrapolation plot to the x-axis. The obtained magnetic dead-layer for (100) and (111) -oriented GGG substrates are 2.88 nm and 5.41 nm, which are comparable to previous reports [37-39]. The saturation magnetization of Bi-YIG films increases as the thickness of the films increases. The increase in saturation magnetization with increase in thickness can be understood by the following ways. Firstly, ferromagnetic thin films are generally deposited with a thin magnetically dead layer over the interface with the substrate. This magnetic dead layer effect is larger in thinner films that leads to the decrease in net magnetization with the decrease in thickness [40,41]. Figure 3 (i) shows the effect of magnetic dead layer region near to the substrate. Secondly, thicker films exhibit the bulk effect of YIG which, in turn, results in increased magnetization.

### 1.3.4 Ferromagnetic resonance study

Figure 4 (a)-(d) shows the FMR absorption spectra of (100) and (111) -oriented films that are labeled with open circle (O) and open triangle (Δ) respectively. FMR experiments were carried out at room temperature. In-plane dc magnetic field was applied parallel to film surface. To find the effective magnetization and Gilbert damping, the FMR linewidth (ΔH) and resonance magnetic field ($H_r$) are calculated using a Lorentzian fit of the FMR absorption spectra measured at $f = 1$ GHz to 12 GHz. Effective magnetization field ($\mu_0 M_{eff}$) were obtained from the fitting of Kittel's in-plane equation (Eq. 3) [42].

$$f = \frac{\gamma}{2\pi} \mu_0 \sqrt{(H_r)(H_r + M_{eff})} \qquad (3),$$

Here, $\mu_0 M_{eff} = \mu_0(M_s - H_{ani})$, anisotropy field $H_{ani} = \frac{2K_1}{\mu_0 M_s}$, and $\gamma$ being the gyromagnetic ratio. Further, the dependence of FMR linewidth on microwave frequency shows a linear variation (Eq. 4) [42] from which the Gilbert damping parameter (α) and FMR linewidth broadening ($\Delta H_0$) were obtained:

$$\mu_0 \Delta H = \mu_0 \Delta H_0 + \frac{4\pi\alpha}{\gamma} f \qquad (4)$$

where, $\Delta H_0$ is the inhomogeneous broadening linewidth and α is the Gilbert damping. Figures 4 (e) and 4 (f) show Kittel and linewidth fitted graphs, respectively. Figure 5 (a)-(d) shows the derived parameters acquired from the FMR study. The estimated Gilbert damping is consistent with data reported for spin-wave propagation [3,22]. The value of α decreases as the thickness of the film increases (Fig. 5 (c)). However, in the instance of Bi-YIG with (111) orientation, there is a substantial increase. This might be attributed qualitatively to the presence of $Bi^{3+}$ ions, which cause strong spin-orbit coupling [43-45] as well as electron scattering inside the lattice when the lattice mismatch (or strain) increases [46]. Our earlier study [7] revealed a clear distribution of $Bi^{3+}$ ions along (111) planes, as well as slightly larger lattice mismatch in Bi-YIG (111). These results explain the larger values of Gilbert damping, $\mu_0 M_{eff}$, and $\Delta H_0$ values in Bi-YIG (111) (Fig. 5). The change in $\mu_0 M_{eff}$ is due to uniaxial in-plane magnetic anisotropy and it is observed from magnetization measurements using $\mu_0 M_{eff} = \mu_0 (M_s - H_{ani})$ [36,47,48]. The enhanced anisotropy field in the lower thickness of Bi-YIG (Fig. 5 (d)) signifies the effect of dead magnetic layer at the interface. The lattice mismatch between films and GGG substrates induces uniaxial in-plane magnetic anisotropy [36,47]. $\Delta H_0$ has a magnitude that is similar to previously published values for the same substrate orientation [7,47]. In conclusion, Bi-YIG with (100) orientation produces the lowest Gilbert damping factor and inhomogeneous broadening linewidth. These are the required optimal parameters for spintronics based devices.

Figure 6 (a) shows the variation of resonance field with polar angle (θ) for the grown 20 nm-150 nm films, $\theta_H$ is the angle measured between applied magnetic field and surface of film (shown in inset of Fig. 4 (a)). The FMR linewidth (ΔH) were extracted from fitting of FMR spectra with Lorentzian absorption functions. From Fig. 6 (a), we observe change in $H_r$ value for 50 nm Bi-YIG film as 0.22 T and 0.27 T for (100) and (111) orientation respectively. Similarly, 0.21 T and 0.31 T change is observed in (100) and (111) orientation respectively for 100 nm Bi-YIG film. We see that $H_r$ increases slightly in case of (111) oriented film by changing the direction of H from 0º to 90º with regard to sample surface (inset of Fig. 4 (a)). The change in $H_r$ decreases with increase in film thickness in case of (100) while it is reversed in case of (111). Figure 6 (b) shows the variation of FMR linewidth with polar angle for 150 nm Bi-YIG film. Maximum FMR linewidth is observed at 90º and it is slightly more as compared with (100) orientation. The enhanced variation of FMR linewidth in (111) oriented samples is generated due to the higher contribution of two-magnon scattering in perpendicular geometry [49]. This can be understood due to the higher anisotropy field in (111) oriented samples (Fig. 5 (d)).

Figure 6 (c) & (e) shows the azimuthal angle (ϕ) variation of $H_r$. Frequency of 5 GHz is used in the measurement. From ϕ variation data (by changing the direction of H from 0º to 360 with regard to sample surface (inset of Fig. 4 (a)). We can see clearly in-plane anisotropy of four-fold in Bi-YIG (100) (Fig. 6 (c)) unlike in Bi-YIG (111) (Fig. 6 (e)). According to crystalline surface symmetry there would be six-fold in-plane anisotropy in case of (111) orientation but we have not observed it, based on previous reports, it can be superseded by a miscut-induced uniaxial anisotropy [33,50]. This reinforces our grown films' single-crystalline nature. The observed change in $H_r$ ($\phi_H$=0 to 45) is 6.6 mT in 50 nm (100), 0.17 mT for 50 nm (111), 6.2 mT in 100 nm (100), 0.17 mT for 100 nm (111)) and 5.1 mT in 150 nm (100). As a result, during in-plane rotation, the higher FMR field change observed along the (100) orientation. The ϕ dependent FMR field data shown in figure 6 (c) were fitted using the following Kittel relation [50]

$$f = \frac{\gamma}{2\pi}\mu_0\sqrt{\begin{array}{l}([H_r\cos(\Phi_H-\Phi_M)+H_c\cos 4(\Phi_M-\Phi_C)+H_u\cos 2(\Phi_M-\Phi_u)]) \times \\ (H_r\cos(\Phi_H-\Phi_M)+M_{eff}+\frac{1}{4}H_c(3+\cos 4(\Phi_M-\Phi_C))+H_u cos^2(\Phi_M-\Phi_u))\end{array}} \quad (5)$$

With respect to the [100] direction of the GGG substrate, in-plane directions of the magnetic field, magnetization, uniaxial, and cubic anisotropies are given by $\Phi_H$, $\Phi_M$, $\Phi_u$ and $\Phi_C$, respectively. $H_u = \frac{2K_u}{\mu_0 M_s}$ and $H_c = \frac{2K_c}{\mu_0 M_s}$ correspond to the uniaxial and cubic anisotropy fields, respectively, with $K_u$ and $K_c$ being the uniaxial and cubic magnetic anisotropy constants, respectively.

Figure 6 (d) shows the obtained uniaxial anisotropy field, cubic anisotropy field and saturation magnetization field for (100) orientation. The obtained saturation magnetization field follows the same pattern as we have obtained from the VSM measurements. The cubic anisotropy field increases and then saturates with the thickness of the film. A large drop in the uniaxial anisotropy field is observed with the thickness of the grown films. We have not got the in-plane angular variation data for the 20 nm thick Bi-YIG sample and may be due to the low thickness of the Bi-YIG, it is not detected by our FMR setup.

## 1.4    Conclusion

In conclusion, we compare the properties of high-quality Bi-YIG thin films of four distinct thicknesses (20 nm, 50 nm, 100 nm, and 150 nm) grown on GGG substrates with orientations of (100) and (111). Pulsed laser deposition was used to synthesize these films. AFM and XRD characterizations reveal that the deposited thin films have smooth surfaces and are phase pure. According to FMR data, the Gilbert damping value decreases with increase in film thickness. This is explained in the context of a dead magnetic layer. The (100) orientation has a lower value of Gilbert damping, indicating that it is the preferable substrate for doped YIG thin films for high frequency application. Bi-YIG on (111) orientation, on the other hand, exhibits anisotropic dominance, which is necessary for magneto-optic devices. The spin-orbit coupled $Bi^{3+}$ ions are responsible for the enhanced Gilbert damping in (111). We have also correlated $\Delta H_0$, anisotropic field, and effective magnetization to the variations in film thickness and substrate orientation. In (100) oriented films, there is unambiguous observation of four-fold in-plane anisotropy. In particular, Bi-YIG grown on (111) GGG substrates yields best result for optimal magnetization dynamics. This is linked to an enhanced magnetic anisotropy. Therefore, proper substrate orientation and thickness are found to be important parameters for growth of Bi-YIG thin film towards high frequency applications.

**Acknowledgments**

This work is supported by the MHRD-IMPRINT grant, DST (SERB, AMT, and PURSE-II) grant of Govt. of India. Ganesh Gurjar acknowledges CSIR, New Delhi for financial support. We acknowledge AIRF, JNU for access of PPMS facility.

**List of figure captions**

**Figure 1:** (a)-(d) X-ray diffraction (XRD) patterns of 20 nm-150 nm Bi-substituted YIG films in (100) and (111) orientations. Insets in (a)-(d) depict the zoomed image of XRD patterns. Variation of lattice constant (e) and (f) lattice mismatch with thickness are shown.

**Figure 2:** (a)-(h) Atomic force microscopy images of 20 nm-150 nm Bi-YIG film in (100) and (111) orientations are shown.

**Figure 3:** (a)-(h) Static magnetization graph of 20 nm-150 nm Bi-substituted YIG (Bi-YIG) films in (100) and (111) orientations. (i) Graph to determine the magnetic dead-layer thickness of Bi-YIG films on (100) and (111)-oriented GGG substrates is depicted (inset shows the variation of saturation magnetization value with the film thickness).

**Figure 4:** (a)-(d) Ferromagnetic resonance (FMR) absorption spectra of 20 nm-150 nm Bi-substituted YIG films with (100) and (111) orientations. Inset in (a) shows the geometry of an applied field angle measured from the sample surface. (e) shows frequency-dependent FMR magnetic field data fitted with Kittel Eq. 3. (f) shows frequency-dependent FMR linewidth data fitted with Eq. 4.

**Figure 5:** Variations of (a) extrinsic linewidth, (b) effective magnetization, (c) Gilbert damping, and (d) magnetic anisotropy with thickness for (100) and (111) oriented Bi-substituted YIG films are depicted.

**Figure 6:** (a) Angular variation of Ferromagnetic resonance (FMR) magnetic field for 20 nm-150 nm Bi-substituted YIG (Bi-YIG) film with (100) and (111) orientations is shown. (b) Angular variation of FMR linewidth of 150 nm thick Bi-YIG film with (100) and (111) orientation is shown. Variations of FMR magnetic field as a function of azimuthal angle ($\phi$) for (c) 50 nm, 100 nm and 150 nm Bi-YIG film with (100) orientation is depicted (d) obtained uniaxial anisotropy field, cubic anisotropy field and saturation magnetization field for (100) orientation. (e) $\phi$ dependent FMR field data for 50 nm and 100 nm Bi-YIG film with (111) orientation is depicted.

**Figure 1**

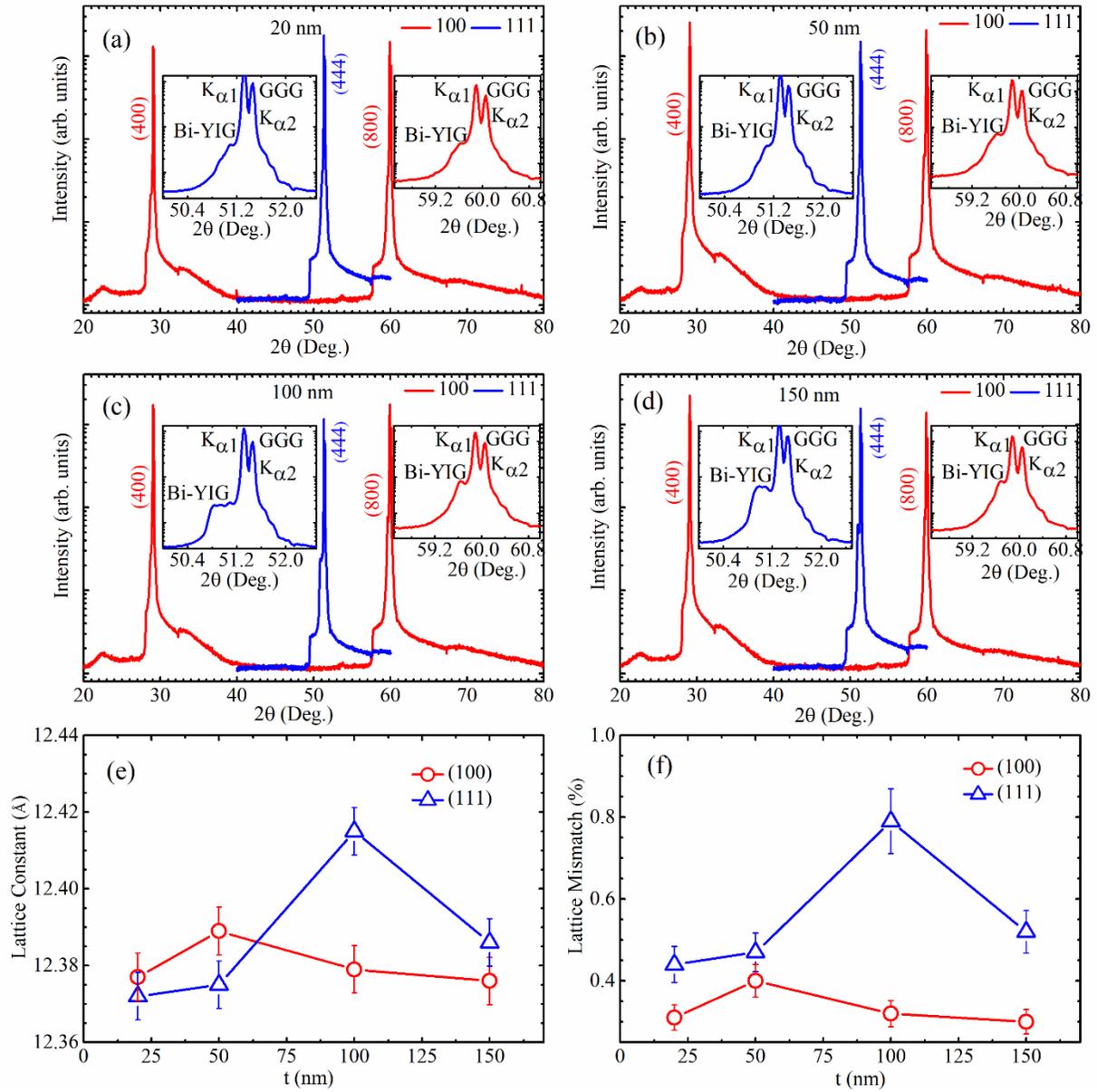

**Figure 2**

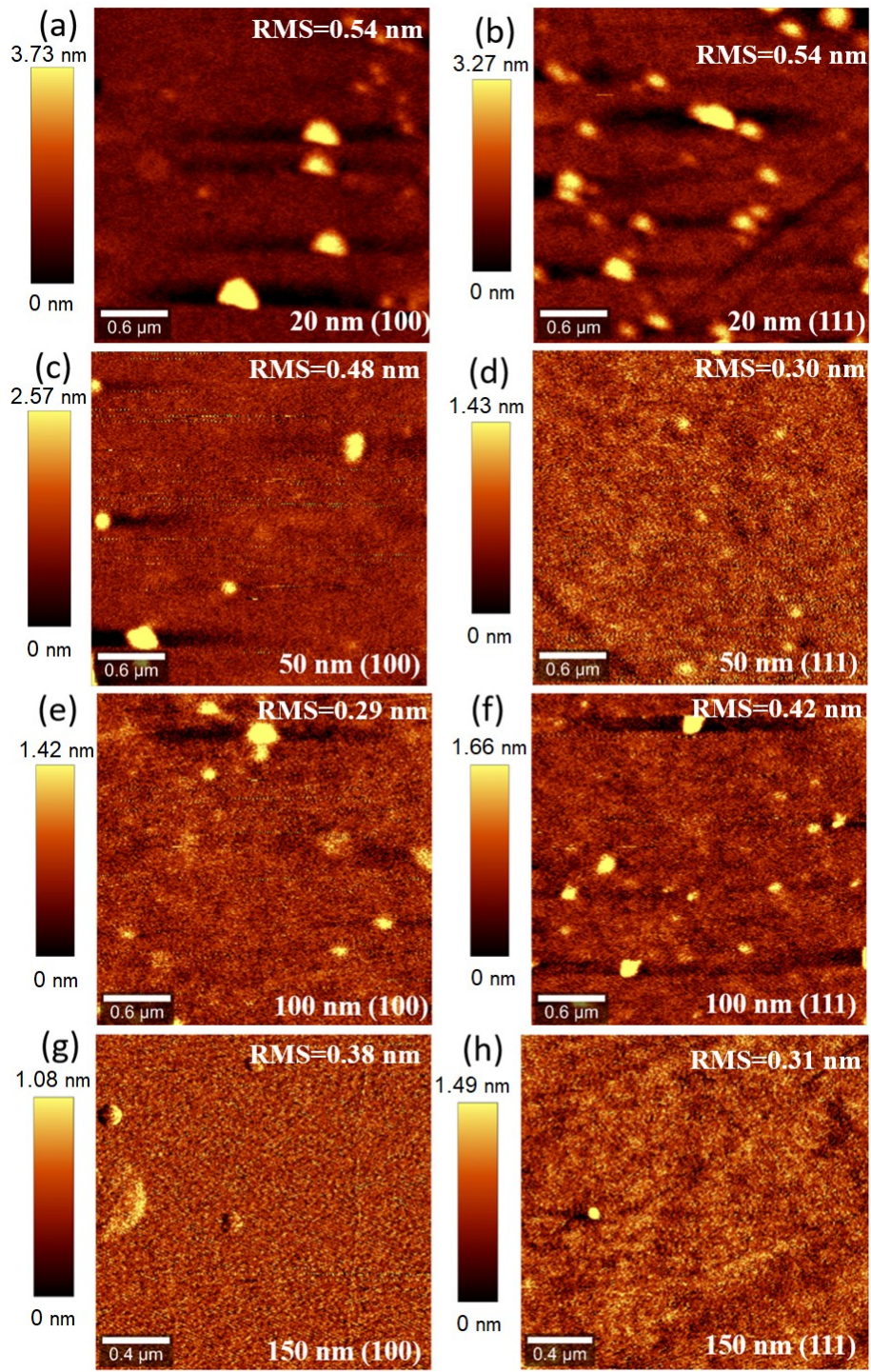

**Figure 3**

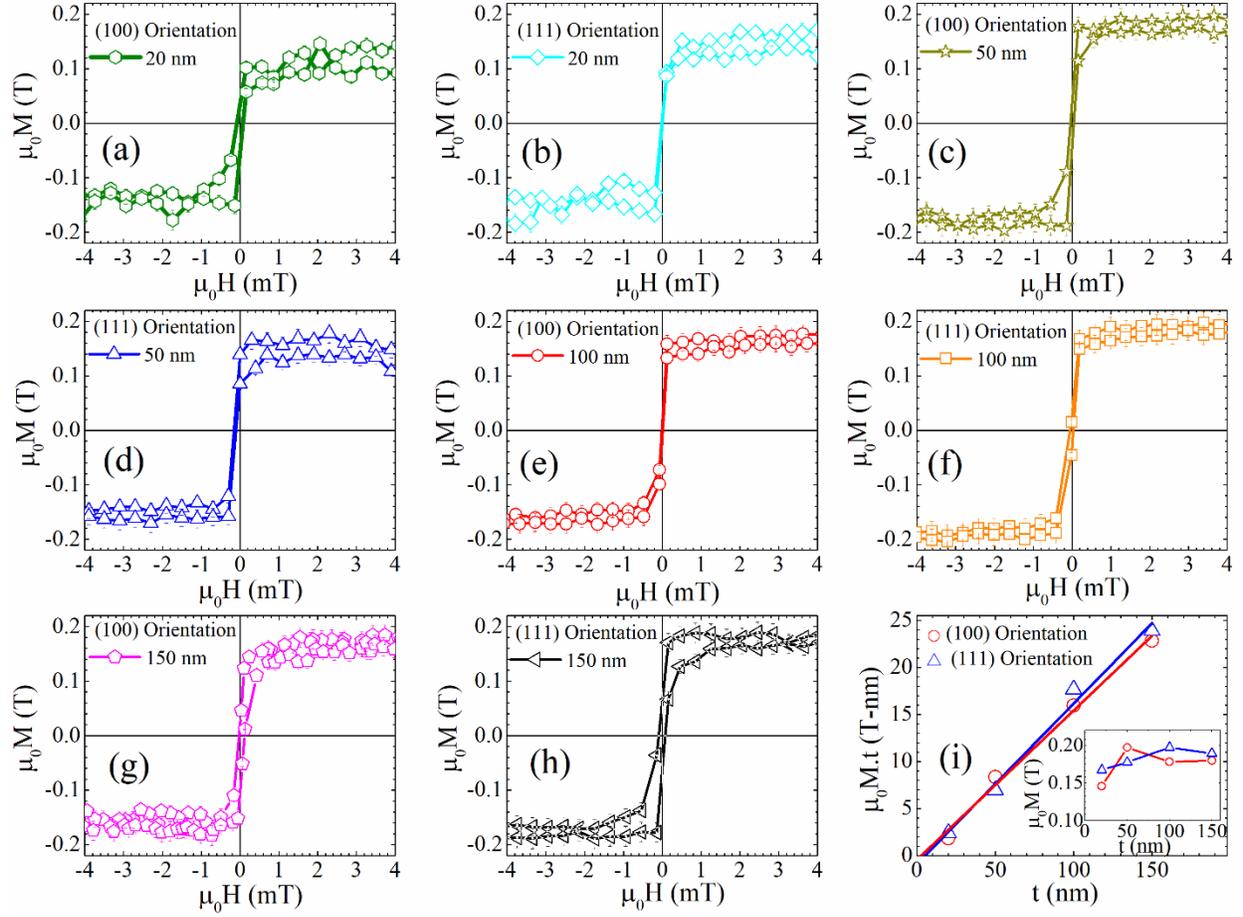

**Figure 4**

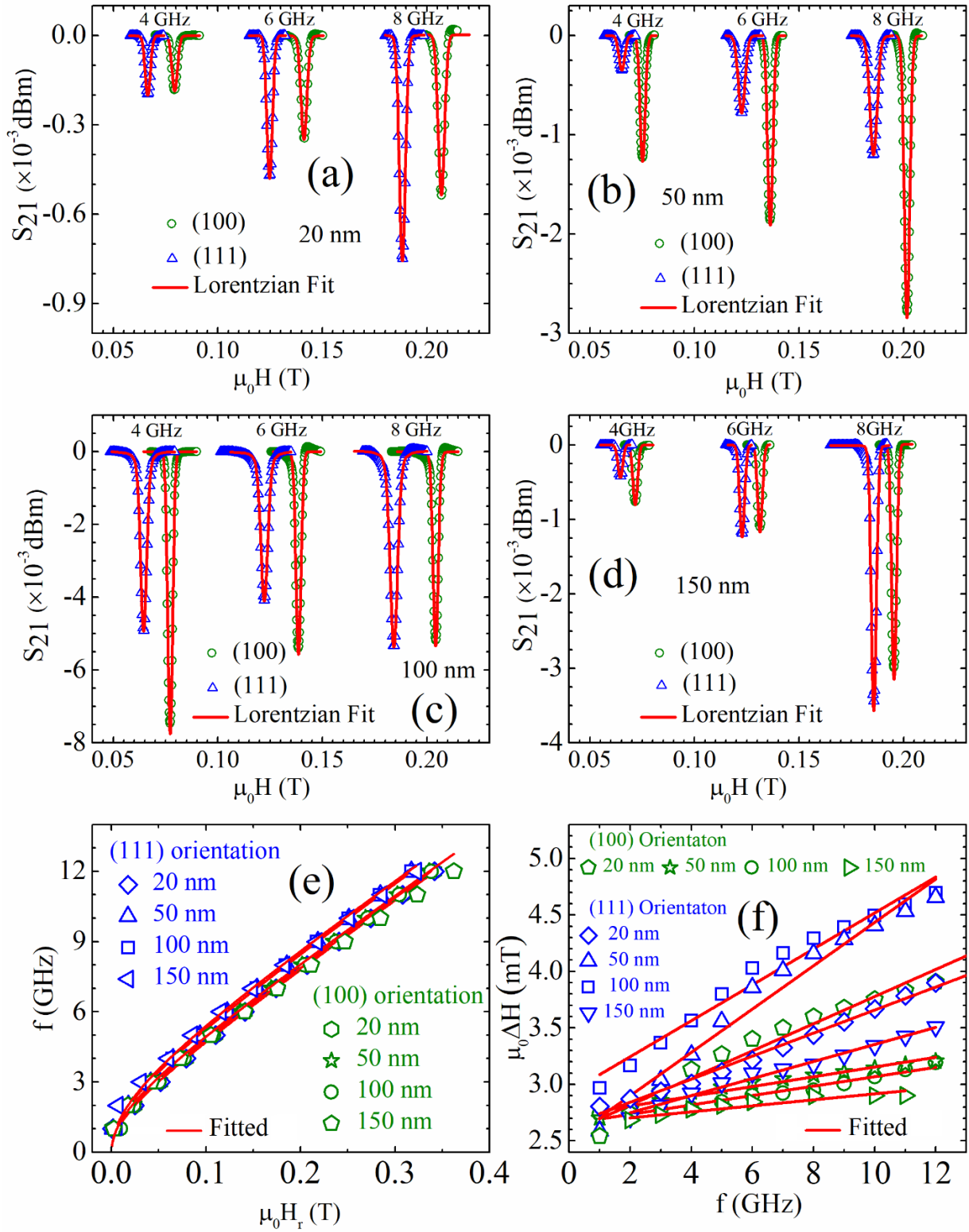

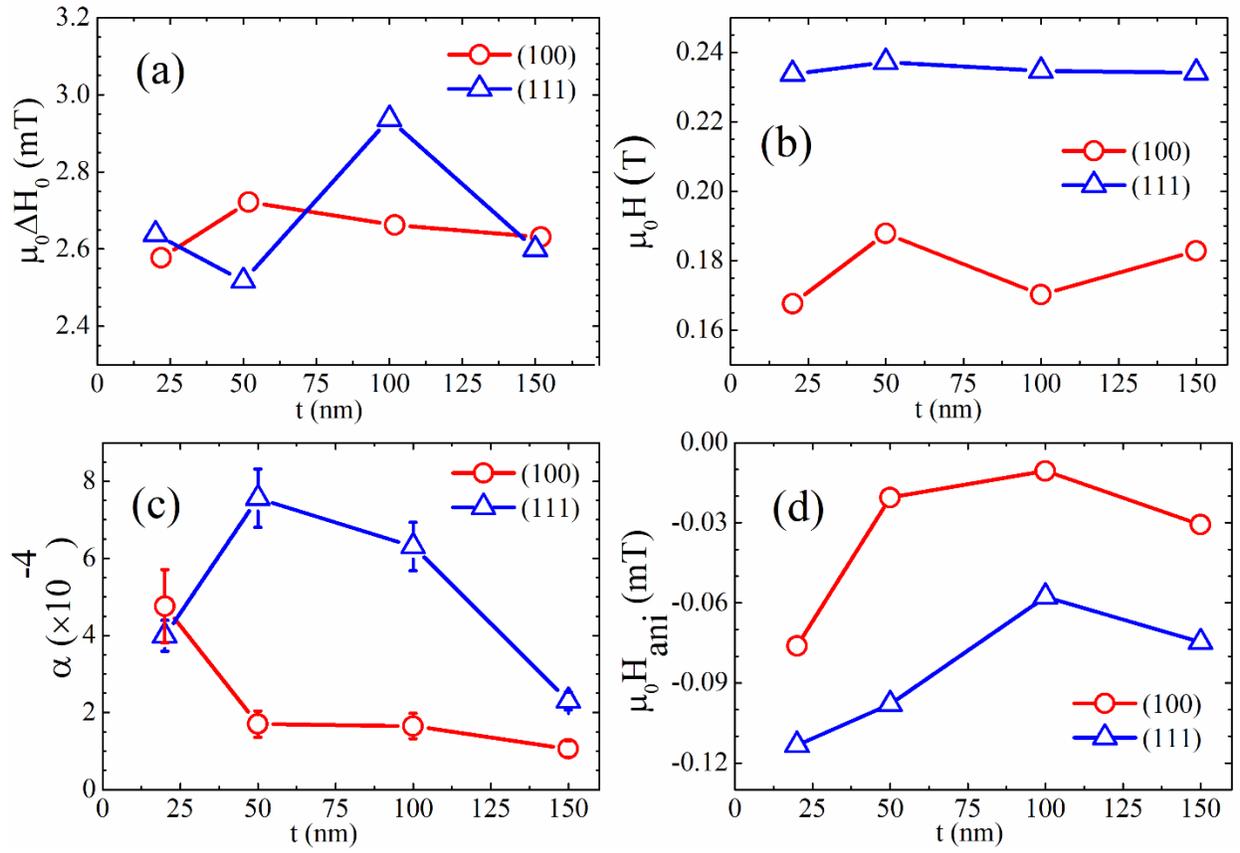

**Figure 6**

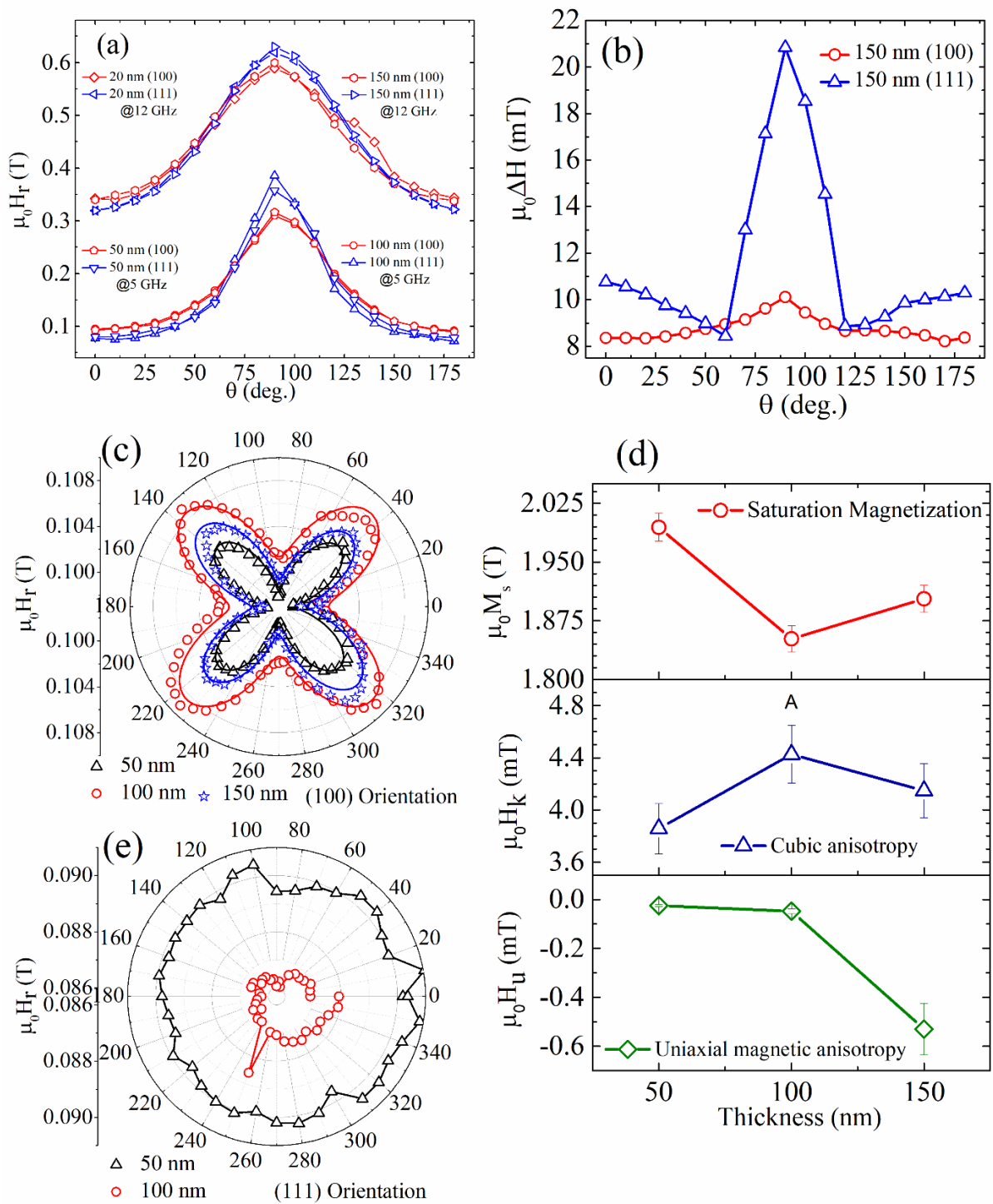